\documentclass[a4paper,11pt]{article}
\usepackage{amsmath,times,amssymb,latexsym,multicol,graphics}
\usepackage{ascmac,fancybox}
\usepackage{bbm}
\usepackage{setspace} 
\usepackage{comment}
\usepackage{authblk}
\usepackage{braket}
\usepackage{genyoungtabtikz}%Young
\newcommand{\be}{\begin{equation}}
\newcommand{\ee}{\end{equation}}
\newcommand{\nn}{\nonumber}
\newcommand{\D}{\Delta}

\newcommand{\G}{\Gamma}
\newcommand{\la}{\lambda}

\newcommand{\p}{\partial}

\usepackage{fullpage}
\begin{document} 
%%%%%%%%%%%%%%%TITLE%%%%%%%%%%%%%%%
\begin{titlepage}
\begin{flushright}
{\small OU-HET 939}
 \\
\end{flushright}

\begin{center}

\vspace{1cm}

\hspace{3mm}{\LARGE \bf Geodesic Witten diagrams with anti-symmetric exchange} \\[3pt] 
%\vspace{1mm}
%{\LARGE \bf }  
 
\vspace{1cm}

\renewcommand\thefootnote{\mbox{$\fnsymbol{footnote}$}}
Kotaro {Tamaoka}

\vspace{5mm}

{\small \sl Department of Physics, Osaka University} \\ 
{\small \sl Toyonaka, Osaka 560-0043, JAPAN}

\vspace{5mm}

{\small{\,k-tamaoka@het.phys.sci.osaka-u.ac.jp}
}

\end{center}

\vspace{5mm}

\noindent
\abstract
We show the AdS/CFT correspondence between the conformal partial wave and the geodesic Witten diagram with anti-symmetric exchange. To this end, we introduce the embedding space formalism for anti-symmetric fields in AdS. Then we prove that the geodesic Witten diagram satisfies the conformal Casimir equation and the appropriate boundary condition. Furthermore, we discuss the connection between the geodesic Witten diagram and the shadow formalism by using the split representation of harmonic function for anti-symmetric fields. We also discuss the 3pt geodesic Witten diagrams and its extension to the mixed-symmetric tensors.
\end{titlepage}
\setcounter{footnote}{0}
\renewcommand\thefootnote{\mbox{\arabic{footnote}}}
%%%%%%%%%%%%%%%TITLE%%%%%%%%%%%%%%%
\tableofcontents
\flushbottom
\section{Introduction}
Conformal Field Theory (CFT) has been studied in various contexts, for example, critical phenomena \cite{Polyakov:1970xd}, exactly solvable models in two dimension \cite{Belavin:1984vu}, and the AdS/CFT correspondence \cite{Maldacena:1997re, Gubser:1998bc, Witten:1998qj}. 
It is well known that the conformal symmetry determines the 2pt and 3pt correlation functions up to constant. In addition, the operator product expansion (OPE) that relates higher point functions to lower ones does converge. 
Therefore, we have universal basis of more than 3pt functions in CFT. These are called the conformal partial waves (CPW). Recently, understanding the structure of CPW have become more and more important. One of the reason is recent resurgence of the conformal bootstrap in more than 2 dimensions \cite{Rattazzi:2008pe, ElShowk:2012ht}. 
For such studies, the explicit form of CPW is necessarily. 

On the other hand, the AdS/CFT correspondence suggests that certain $d$-dimensional CFT give the plausible non-perturbative definition of quantum gravity in asymptotically AdS${}_{d+1}$. 
If so, describing quantum gravity in the language of CFT is quite reasonable, at least perturbation on $1/N$, hence CPW can be fundamental building blocks. Recently, the bulk dual of CPW has turned out to be geodesic Witten diagrams (GWD) \cite{Hijano:2015zsa}, that is, Witten diagrams whose integral domains are restricted on the geodesics. Note that we can apply bootstrap techniques to CFT expected to have the bulk dual \cite{Heemskerk:2009pn, ElShowk:2011ag}. 
  
CPW is solution of the conformal Casimir equation \cite{Dolan:2003hv}. Therefore GWD should satisfy this property. There is another method to solve the conformal Casimir equations which is called the shadow formalism \cite{Ferrara:1972xe,Ferrara:1972uq,Ferrara:1972ay,Ferrara:1973vz,Ferrara:1974nf,SimmonsDuffin:2012uy}. The advantages of GWD, compared to the shadow formalism, are (1) the direct connection to the bulk theories, and (2) no shadow contribution. We will address these properties in section \ref{sec:gwd}. 
  
For such quest, studying CPW with mixed-symmetric tensors is important. Even when we only consider the correlation function of the symmetric-traceless fields, internal operator can be other than the symmetric-traceless ones, namely mixed-symmetric fields. Such CPW go on the stage, for example, when bootstrapping the correlation function for spinning operators, and studying the higher spin theory in AdS \cite{Vasiliev:2004cm}. 

From the above motivations, we should study GWD with mixed-symmetric tensors. 
GWD was first developed by \cite{Hijano:2015zsa} for CPW with external scalars. Extension to the external symmetric-traceless fields was studied by \cite{Nishida:2016vds,Castro:2017hpx,Dyer:2017zef,Sleight:2017fpc, Chen:2017yia}. See also related work which includes the GWD with interfaces \cite{Rastelli:2017ecj}. 
In these studies, the internal operator was restricted on the scalar or symmetric-traceless field. 

In this paper, we discuss the anti-symmetric ($p$-form) exchange especially. This is because, in this case, the bulk-bulk propagators has been known explicitly \cite{Bena:1999be, Bena:1999py, Naqvi:1999va}. As an explicit example, we display the correspondence between GWD and CPW for the two scalar and two vector fields with $2$-form exchange. In addition, we compute the 3pt diagram including a mixed-symmetric tensor (described by Young diagrams with the hook) whose bulk-boundary propagator can be derived from conformal symmetry. On these accounts, first we extend the embedding formalism in AdS \cite{Costa:2014kfa} to the anti-symmetric tensors (indices). Moreover, we discuss the split representation of the harmonic function for anti-symmetric fields so that we can see the connection between GWD and the shadow formalism. 

This paper is organized as follows. In section \ref{sec:emb}, we develop the embedding formalism for anti-symmetric indices in AdS space by introducing auxiliary Grassmann odd fields. We also comment on the bulk-boundary propagator for mixed-symmetric tensors. In section \ref{sec:gwd3}, we compute some 3pt GWD with mixed-symmetric fields and discuss the choice of 3pt interactions and geodesics. In section \ref{sec:gwd}, we show the correspondence between CPW and GWD with anti-symmetric exchange and its relation to the shadow formalism. In section \ref{sec:summary}, we summarize this work and discuss the future directions. 

\section{Embedding formalism in AdS${}_{d+1}$}\label{sec:emb}
In this section, we develop the embedding formalism for anti-symmetric fields in AdS. 
Since AdS${}_{d+1}$ isometry $SO(d+1,1)$ can be regarded as Lorentz symmetry in $\mathbb{R}^{d+1,1}$, we can naturally embed AdS${}_{d+1}$ fields into $\mathbb{R}^{d+1,1}$. The formalism for symmetric-traceless fields was studied intensively in \cite{Costa:2014kfa} (see also earlier work in \cite{Joung:2011ww}).  The embedding formalism was also developed for CFT rather earlier \cite{Costa:2011mg,Costa:2011dw, Costa:2014rya,Dirac:1936fq,Mack:1969rr,Boulware:1970ty,Ferrara:1973eg,Ferrara:1973yt,Cornalba:2009ax,Weinberg:2010fx}. 
\subsection{Anti-symmetric tensors in embedding space}
%%%
\if(
\begin{figure}[tbp]
\begin{center}
\resizebox{80mm}{!}{\includegraphics{nullandads.eps}}
\caption{Euclidian AdS${}_{d+1}$ (red hyperboloid) and its $d$-dimensional conformal boundary (blue light cone $P^2=0, P^0>0$) in the embedding space $\mathbb{R}^{d+1,1}$. The blue light ray represents the identification of the boundary points $P^A\sim\la P^A$. The black hyperbolic curve shows the flat section for the boundary (the Poincar\'e section). }\label{fig:nullandads}
\end{center}
\end{figure} 
)\fi
%%%
We use the embedding definition of (Euclidian) AdS${}_{d+1}$ and its conformal boundary $\mathbb{R}^{d}$. Let us consider $\mathbb{R}^{d+1,1}$ and its sub-manifolds
\begin{align}
\textrm{AdS}_{d+1}:\;& \;X^2=-1, \;X^0>0, \\
\mathbb{R}^{d}:\;&\;P^2=0, \;P^A\sim \la P^A\hspace{3mm}(\la\in\mathbb{R}). 
\end{align}
These sub-manifolds indeed represent the AdS${}_{d+1}$ and its conformal boundary $\mathbb{R}^{d}$. Especially, 
\be
X^A=(X^+,X^-,X^a)=\dfrac{1}{z}\left(1,z^2+x^2,x^a\right),
\ee
and
\be
P^A=(P^+,P^-,P^a)=\left(1,x^2,x^a\right),
\ee
describe the Poincar\'e patch and its boundary. %(see figure \ref{fig:nullandads}). 
Here we used the light cone metric
\be
A\cdot B=\eta_{AB}A^AB^B=-\dfrac{1}{2}\left(A^+B^-+A^-B^+\right)+\delta_{ab}A^aB^b \hspace{3mm}(\textrm{for all vector $A, B$ in}\;\mathbb{R}^{d+1,1}). 
\ee

Let us consider the anti-symmetric $p$-tensor ($p$-form) field $F_{A_1\cdots A_p}(X)$ in the embedding AdS space. This field can be pulled back to the original field $f_{\mu_1\cdots \mu_p}(x)$ in AdS via
\be
f_{\mu_1\cdots \mu_p}(x)=\dfrac{\p X^{A_1}}{\p x^{\mu_1}}\cdots\dfrac{\p X^{A_p}}{\p x^{\mu_p}}F_{A_1\cdots A_p}(X), 
\ee
where $x^\mu=(z,x^a)$ is the original AdS coordinates. 
We impose the transverse condition
\be
X^{A_1}F_{A_1\cdots A_p}(X)=0, 
\ee
in such a way that the number of independent components of $F_{A_1\cdots A_p}$ corresponds to the one of $f_{\mu_1\cdots \mu_p}$. This condition projects out the unphysical components, such as
\be
F_{A_1\cdots A_p}(X)=X_{[A_1}\widetilde{F}_{A_2\cdots A_p]}. \label{eq:transverse}
\ee
Here $\widetilde{F}_{A_2\cdots A_p}$ is an anti-symmetric tensor which also satisfies the similar condition as \eqref{eq:transverse}. 
Thus, there are in total $\binom{d+1}{p-1}$ constraints so that the number of components of $F_{A_1\cdots A_p}(X)$ can accord with the original $f_{\mu_1\cdots \mu_p}(x)$. In the embedding space, it is very useful to introduce the auxiliary field to contract the indices
\be
F(X,\Theta)\equiv\Theta^{A_1}\cdots\Theta^{A_p}F_{A_1\cdots A_p}(X), 
\ee
where $\Theta$ is a Grassmann odd field. We can restrict $\Theta\cdot X=0$ without loss of any information ($\Theta\cdot\Theta=0$ is automatically satisfied). This auxiliary field encode $p$-form field into the polynomial of $\Theta$ with $p$-degrees. 
In order to go back the components language, we introduce the differential operator
\be
K^{\Theta}_A=\dfrac{\p}{\p\Theta^A}+X_A\left(X\cdot\dfrac{\p}{\p\Theta}\right). \label{eq:K}
\ee
This operator will decode a given polynomial for $\Theta$ into an anti-symmetric tensor. Note that $K_A$ is interior with respect to the submanifold $X^2+1=(\Theta\cdot X)=0$. Throughout this paper, we use left derivative for Grassmann variables. In addition, it satisfies\footnote{We use the anti-symmetrization $[\cdots]$ with strength $1$.}
\begin{subequations}
\begin{align}
K^{\Theta}_AK^{\Theta}_B=-K^{\Theta}_BK^{\Theta}_A\hspace{1cm}&(\textrm{anti-symmetric}), \\
X^AK^{\Theta}_A=0 \hspace{1cm}&(\textrm{transverse}), \\
K^{\Theta}_{A_1}\cdots K^{\Theta}_{A_p}\Theta^{B_1}\cdots\Theta^{B_p}=(-1)^{\frac{1}{2}p(p-1)}\,G^{B_1}_{[A_1}\cdots G^{B_p}_{A_p]}\hspace{1cm}&(\textrm{projection}). 
\end{align}
\end{subequations}
Here $G_{AB}$ is the induced metric
\be
G_{AB}=\eta_{AB}+X_AX_B.
\ee
For general tensor fields, the covariant derivative $\nabla_A$ is naturally defined as
\be
\nabla_AF_{A_1\cdots A_p}(X)=G^B_AG^{B_1}_{A_1}\cdots G^{B_p}_{A_p}\dfrac{\p}{\p X^B}\,F_{B_1\cdots B_p}(X), 
\ee
which preserves the transverse condition $X^A\nabla_A=0$. 
When we consider the polynomial of $\Theta$, the counterpart of the above covariant derivative is
\be
\nabla_AF(X,\Theta)=\left[\dfrac{\p}{\p X^A}+X_A\left(X\cdot\dfrac{\p}{\p X}\right)+\Theta_A\left(X\cdot\dfrac{\p}{\p\Theta}\right)\right]F(X,\Theta). 
\ee  
The last term of r.h.s. is necessary to stay in the submanifold $(\Theta\cdot X)=0$. These definitions become equivalent to $\nabla_\mu f_{\mu_1\cdots\mu_p}$ after projecting to the original AdS space. 

The $SO(d+1,1)$ generator for the anti-symmetric fields is
\be
L_{AB}=X_A\dfrac{\p}{\p X^B}-X_B\dfrac{\p}{\p X^A}+\Theta_A\dfrac{\p}{\p\Theta^B}-\Theta_B\dfrac{\p}{\p\Theta^A}. 
\ee
For the $p$-form polynomial $F(X,\Theta)$, one can see
\be
-\dfrac{1}{2}L_{AB}L^{AB}F(X,\Theta)=[\nabla^2+p(d+1-p)]F(X,\Theta), 
\ee
where $\nabla^2$ is the Laplacian for $p$-form in the embedding space: 
\be
\nabla^2F(X,\Theta)=\left[G^{AB}\dfrac{\p}{\p X^A}\dfrac{\p}{\p X^B}+(d+1)\left(X\cdot\dfrac{\p}{\p X}\right)-\left(\Theta\cdot\dfrac{\p}{\p\Theta}\right)\right]F(X,\Theta). 
\ee
\subsection{Anti-symmetric tensor propagators in embedding space}
Let us consider anti-symmetric tensor ($p$-form) fields. The equation of motion for a massive\footnote{We only consider massive fields for simplicity. } $p$-form with source field $j_\mu$ is
\be
\dfrac{1}{p!}\nabla^{\mu}\p_{[\mu}A_{\mu_1\mu_2\cdots \mu_p]}-m^2A_{\mu_1\mu_2\cdots \mu_p}=j_{\mu_1\cdots \mu_p}, \label{eq:p-formEOM}
\ee
Then the bulk-bulk propagator satisfies
\be
\dfrac{1}{p!}\nabla_1^{\mu}\p_{[\mu}G_{\mu_1\cdots \mu_p]}{}^{\nu_1\cdots \nu_p}(x_1,x_2)-m^2G_{\mu_1\cdots \mu_p}{}^{\nu_1\cdots \nu_p}(x_1,x_2)=\delta(x_1,x_2)g_{[\mu_1\cdots}^{\nu_1}\cdots g_{\mu_p]}^{\nu_p}. 
\ee
If the source term is absent, by using the Riemann curvature in AdS${}_{d+1}$
\be
R^{\mu\nu\rho\sigma}=-(g^{\mu\rho}g^{\nu\sigma}-g^{\mu\sigma}g^{\nu\rho}), 
\ee
one can show that \eqref{eq:p-formEOM} is equivalent to
\begin{align}
[\nabla^2-\D(\D-d)+p]A_{\mu_1\mu_2\cdots \mu_p}&=0, \\
\nabla^{\mu_1}A_{\mu_1\mu_2\cdots \mu_p}&=0. 
\end{align}
Here we used the relation between mass and scaling dimension $\D$, $m^2=\D(\D-d)+p(d-p)$. Note, however, the similar modification for bulk-bulk propagator is justified only when $x_1\neq x_2$. 
\subsubsection{Bulk-bulk propagator}
The tensor structure of bulk-bulk propagator is fixed by AdS isometry. Namely, we need to have the polynomials of degree $p$ in both $\Theta_1$ and $\Theta_2$. 
\begin{align}
G^{\D,[p]}_{bb}(X_1,\Theta_1;X_2,\Theta_2)&=(\Theta_1\cdot\Theta_2)^pg_0(U)+(\Theta_1\cdot\Theta_2)^{p-1}(\Theta_1\cdot X_2)(\Theta_2\cdot X_1)g_1(U). \label{eq:pGbb}
\end{align}
Note that $(\Theta_i\cdot X_j)^2=0$. Bulk-bulk propagator for the $p$-form was studied in \cite{Bena:1999be,Bena:1999py,Naqvi:1999va}.  We displayed only results in the embedding space\footnote{The expression \eqref{eq:g0} and \eqref{eq:g1} may look like different from the literatures, for example, \cite{Naqvi:1999va}. However, one can check that this expression is equivalent to one of \cite{Naqvi:1999va}. We note useful identities for the hypergeometric function to check it. 
\begin{align}
z\dfrac{d}{dz}{}_2F_1(\alpha,\beta,\gamma,z)&=\alpha\big({}_2F_1(\alpha+1,\beta,\gamma,z)-{}_2F_1(\alpha,\beta,\gamma,z)\big), \nn\\
z(1-z)\dfrac{d}{dz}{}_2F_1(\alpha,\beta,\gamma,z)&=(\alpha-\gamma+\beta z){}_2F_1(\alpha,\beta,\gamma,z)+(\gamma-\alpha){}_2F_1(\alpha-1,\beta,\gamma,z).\nn
\end{align}
When $p=1$, these are consistent with (44)-(47) of \cite{Costa:2014kfa}.}, 
\begin{subequations}
\begin{align}
g_0(U)&=(d-\D)F_1(U)-p\dfrac{1+U}{U}F_2(U), \label{eq:g0}\\
g_1(U)&=p\left[\dfrac{1+U}{U(U+2)}(d-\D)F_1(U)-\dfrac{pU(U+2)+d+1}{U^2(U+2)}F_2(U)\right], \label{eq:g1}
\end{align}
\end{subequations}
where
\begin{subequations}
\begin{align}
F_1(U)&=\mathcal{N}(2U^{-1})^\D\,{}_2F_1\left(\D,\D-\frac{d}{2}+\dfrac{1}{2},2\D-d+1,-2U^{-1}\right), \\
F_2(U)&=\mathcal{N}(2U^{-1})^\D\,{}_2F_1\left(\D+1,\D-\frac{d}{2}+\dfrac{1}{2},2\D-d+1,-2U^{-1}\right), \\
\mathcal{N}&=\dfrac{\Gamma(\D+1)}{2\pi^{\frac{d}{2}}(d-p-\D)(\D-p)\Gamma\left(\D+1-\frac{d}{2}\right)}.
\end{align}
\end{subequations}
Here $U$ is the chordal distance defined as $U=-1-X_1\cdot X_2$. 
By using the above ingredients, we have obtained
\begin{align}
-\dfrac{1}{2}L^2_1 G^{\D,[p]}_{bb}(X_1, \Theta_1; X_2, \Theta_2)&=[\nabla^2_1+p(d+1-p)]G^{\D,[p]}_{bb}(X_1, \Theta_1; X_2, \Theta_2)\nn\\
&=m^2G^{\D,[p]}_{bb}(X_1, \Theta_1; X_2, \Theta_2)\hspace{5mm}(\textrm{If} \;X_1\neq X_2)\nn\\
&=C_{\D,[p]}G^{\D,[p]}_{bb}(X_1, \Theta_1; X_2, \Theta_2) \hspace{5mm}(\textrm{If} \;X_1\neq X_2), \label{eq:asmgwd}
\end{align}
where $C_{\D,[p]}=\D(\D-d)+p(d-p)$ is the quadratic conformal Casimir for $p$-form. This is just the key identity for the geodesic Witten diagrams with the $p$-form exchange. It should be again stressed that the divergence of the propagator becomes subtle when $X_1=X_2$. Therefore the second equality is valid only when two points $X_1$ and $X_2$ are separated although it is enough for our geodesic integrals. 

\subsubsection{Bulk-boundary propagator}
The tensor structure of bulk-boundary propagators are completely fixed by the conformal symmetry. This is also the case for $p$-form propagator $G^{\D,[p]}_{b\p}(X,\Theta;P,\theta)$. It satisfies
\be
G^{\D,[p]}_{b\p}(X,\alpha_1\Theta+\beta_1 X;\la P,\alpha_2\theta+\beta_2 P)=\la^{-\D}(\alpha_1\alpha_2)^{p}G^{\D,[p]}_{b\p}(X,\Theta;P,\theta), 
\ee
and we have unique tensor structure which satisfies the above condition: 
\begin{align}
G^{\D,[p]}_{b\p}(X,\Theta;P,\theta)&=\dfrac{\mathcal{C}_{\Delta,[p]}}{(-2X\cdot P)^{\D+p}}[(\theta\cdot\Theta)(P\cdot X)-(X \cdot \theta)(P\cdot \Theta)]^p, 
\end{align}
where $\mathcal{C}_{\Delta,[p]}$ is the overall constant to be determined below. 

Besides the kinematical derivation, let us see the boundary limit of the bulk-bulk propagator which fixes the overall constant. 
\begin{align}
\lim_{\lambda\rightarrow\infty}\la^\D\,G^{\D,[p]}_{bb}(X,\Theta; \la P+\mathcal{O}(\la^{-1}),\theta)&=\dfrac{\mathcal{C}_{\Delta,[p]}}{(-2X\cdot P)^{\D}}\left[(\theta\cdot\Theta)^p-p\,(\theta\cdot\Theta)^{p-1}\dfrac{(X\cdot\theta)(P\cdot\Theta)}{X\cdot P}\right]\nn\\
&=\dfrac{\mathcal{C}_{\Delta,[p]}}{(-2X\cdot P)^{\D+p}}[(\theta\cdot\Theta)(P\cdot X)-(X \cdot \theta)(P\cdot \Theta)]^p,\nn\\
\mathcal{C}_{\Delta,[p]}&=\dfrac{\G(\D+1)}{2\pi^{\frac{d}{2}}(\D-p)\Gamma\left(\D+1-\frac{d}{2}\right)}. 
\end{align}

\subsection{Bulk-boundary propagator for the mixed-symmetric tensor}\label{subsec:H}
Before closing this section, we comment on the extension to the mixed-symmetric tensors in AdS. We mention only the bulk-boundary propagators and not argue the bulk-bulk propagators. As like \eqref{eq:transverse}, we should impose the transverse condition to the mixed-symmetric field $H$ in order to reduce the degrees of freedom.
Following the arguments on \cite{Costa:2014rya} that discussed the mixed-symmetric fields in CFT, we can find the bulk-boundary propagators for the corresponding ones. This is simply because the bulk-boundary propagators should be compatible with the conformal symmetry. 
The resulting tensor structure is just the combination of $(X\cdot P)\,\eta_{AB}-X_AP_B$ with the appropriate mixed-symmetrization. So quick derivation in the embedding space is just preparing the 2pt CFT correlators and replacing the one side of CFT vectors for the AdS ones. 

We illustrate the simplest example, the irreducible representation corresponded to the smallest hook diagram $\Yboxdim{5pt}\yng(2,1)$\,, since we will use it in next section. The bulk-boundary propagator is
\begin{align}
G^{\D,\,\Yboxdim{3pt}\yng(2,1)}_{b\p}(X, W, \Theta; P, Z, \theta)&=\dfrac{(Z\cdot\p_\theta)(W\cdot \p_\Theta)}{(-2X\cdot P)^{\D+3}}\big[C_1\big(H^{(\Theta,\theta)}_{b\p}H^{(\Theta,\theta)}_{b\p}H^{(W,Z)}_{b\p}\big)+C_2\big(H^{(\Theta,Z)}_{b\p}H^{(\Theta,\theta)}_{b\p}H^{(W,\theta)}_{b\p}\big)\big]\nn\\
&=\dfrac{\mathcal{C}_{\Delta,\,\Yboxdim{3pt}\yng(2,1)}}{(-2X\cdot P)^{\D+3}}\big[H^{(\Theta,\theta)}_{b\p}H^{(W,Z)}_{b\p}-H^{(\Theta,Z)}_{b\p}H^{(W,\theta)}_{b\p}\big]H^{(W,Z)}_{b\p},\label{eq:bph} 
\end{align}
where we defined
\be
H^{(A,B)}_{b\p}=-2[(X\cdot P)(A\cdot B)-(A\cdot P)(B\cdot X)]. 
\ee
Here we employed $W$ and $Z$ as Grassmann even auxiliary fields for the bulk and the boundary respectively. In the first line of \eqref{eq:bph}, we assigned $\Theta (\theta)$ for the left column of the $\Yboxdim{5pt}\yng(2,1)$\,\,and $W (Z)$ for the right column (a box) of one. The two terms in the bracket are all possible tensor structures. The anti-symmetrization has been done in the second line of \eqref{eq:bph}. Notice that the resulting tensor structure is unique. Remaining symmetrization and removing the trace part should be done when we move to the component expression. So we need the appropriate differential operators to do it. The systematic way of finding such operators is involved in general but can be seen in \cite{Costa:2016hju}. 
\section{Three point diagrams on the geodesics}\label{sec:gwd3}
In this section, we discuss the 3pt GWD and related 3pt functions in CFT. 
It is well known that 3pt Witten diagram is equivalent to the 3pt functions in CFT. Since 3pt functions are fixed by conformal symmetry, 3pt GWD also reproduce its functional form (though we cannot obtain the OPE coefficient directly, of course). %We also see the subtlety of interactions. 
\subsection{Three point diagrams with $p$-form} \label{subsec:paf}
First, we consider 3pt diagram for scalar-vector-($2$-form). We will use it for the later discussion in section \ref{subsec:shadow}. For a moment, we assign 3pt interaction
\be
S_{int}=\la_{\phi AF} \int_{\textrm{AdS}} dX (\p_I\phi_1)A_{2J}F^{IJ}_3. \label{eq:3pint_paf}
\ee
Here $\phi_1 ,A_{2J}$ and $F_{3}^{IJ}$ are supposed to be the bulk dual operator of CFT, $\mathcal{O}_1, \mathcal{J}_{2I}$, and $\mathcal{F}_{3}^{IJ}$ respectively. Our (tree-level) amplitude is then
\be
\mathcal{A}_3=\la_{\phi AF}\int_{\textrm{AdS}} dX \left[(\p_IG^{\D_1,0}_{b\p}(X;P_1))[G^{\D_2,1}_{b\p}(X;P_2,Z_2)]_J[G^{\D_3,[2]}_{b\p}(X;P_3,\theta_3)]^{IJ}\right]. 
\ee
Here each bulk-boundary propagator was defined as follows\footnote{For the symmetric-traceless case, we basically follow the notation of \cite{Costa:2014kfa}.},
\begin{subequations}
\begin{align}
G^{\D_1,0}_{b\p}(X;P_1)&=\mathcal{C}_{\D_1,0}\dfrac{1}{(-2X\cdot P_1)^{\D_1}}, \\
[G^{\D_2,1}_{b\p}(X;P_2,Z_2)]_J&=\mathcal{C}_{\D_2,1}\dfrac{1}{(-2X\cdot P_2)^{\D_2+1}}\left[(-2P_2\cdot X)Z_{2J}+2P_{2J}(Z_2\cdot X)\right], \\
[G^{\D_3,[2]}_{b\p}(X;P_3,\theta_3)]_{IJ}&=\dfrac{\mathcal{C}_{\D_3,[2]}}{2}\dfrac{1}{(-2X\cdot P_3)^{\D_3+2}}(G^{[A_1}_IG^{A_2]}_J)\prod^2_{i=1}\left[(-2P_3\cdot X)\theta_{3A_i}+2P_{3A_i}(X\cdot \theta_3)\right].
\end{align}
\end{subequations}
Although this integral can be done explicitly in principle, from here we discuss the geodesic integrals instead of the entire bulk integrals. We will denote 3pt GWD as $\mathcal{W}_3(\gamma_{ij})$, where $\gamma_{ij}$ represents the geodesics anchored on the boundary points $P_i$ and $P_j$. By using the embedding coordinates, this geodesics can be expressed as
\be
\gamma_{ij}\;:\;X^A(\la)=\dfrac{e^{-\la}P_i+e^{\la}P_j}{\sqrt{-2P_i\cdot P_j}}, 
\ee
where $\la$ is the proper time of the geodesics. Our geodesic integral $\mathcal{W}_3(\gamma_{ij})$ is then
\begin{align}
\mathcal{W}_3(\gamma_{ij})&=\la_{\phi AF}\int_{\gamma_{ij}} d\la\, \left[(\p_IG^{\D_1,0}_{b\p}(X(\la);P_1))[G^{\D_2,1}_{b\p}(X(\la);P_2,Z_2)]_J[G^{\D_3,[2]}_{b\p}(X(\la);P_3,\theta_3)]^{IJ}\right]\nn\\
&=\int_{\gamma_{ij}} d\la\,\dfrac{T(P_i,Z_2,\theta_3)}{(-2X(\la)\cdot P_1)^{\D_1+1}(-2X(\la)\cdot P_2)^{\D_2+1}(-2X(\la)\cdot P_3)^{\D_3+2}}, 
\end{align}
where
\begin{align}
T(P_i,Z_2,\theta_3)&=\la_{\phi AF}\D_1\mathcal{C}_{\D_1,0}\,\mathcal{C}_{\D_2,1}\,\mathcal{C}_{\D_3,[2]}[(P_1\cdot \theta_3)(2X(\la)\cdot P_3)-(P_1\cdot P_3)(2X(\la)\cdot \theta_3)]\nn\\
&\times[(-2P_2\cdot X(\la))Z_{2A}+2P_{2A}(Z_2\cdot X(\la))][\theta_{3}^A(2X(\la)\cdot P_3)-P_{3}^A(2X(\la)\cdot \theta_3)], 
\end{align}
is tensor polynomial part of our amplitude. 
\begin{figure}[tbp]
\begin{center}
\resizebox{140mm}{!}{\includegraphics{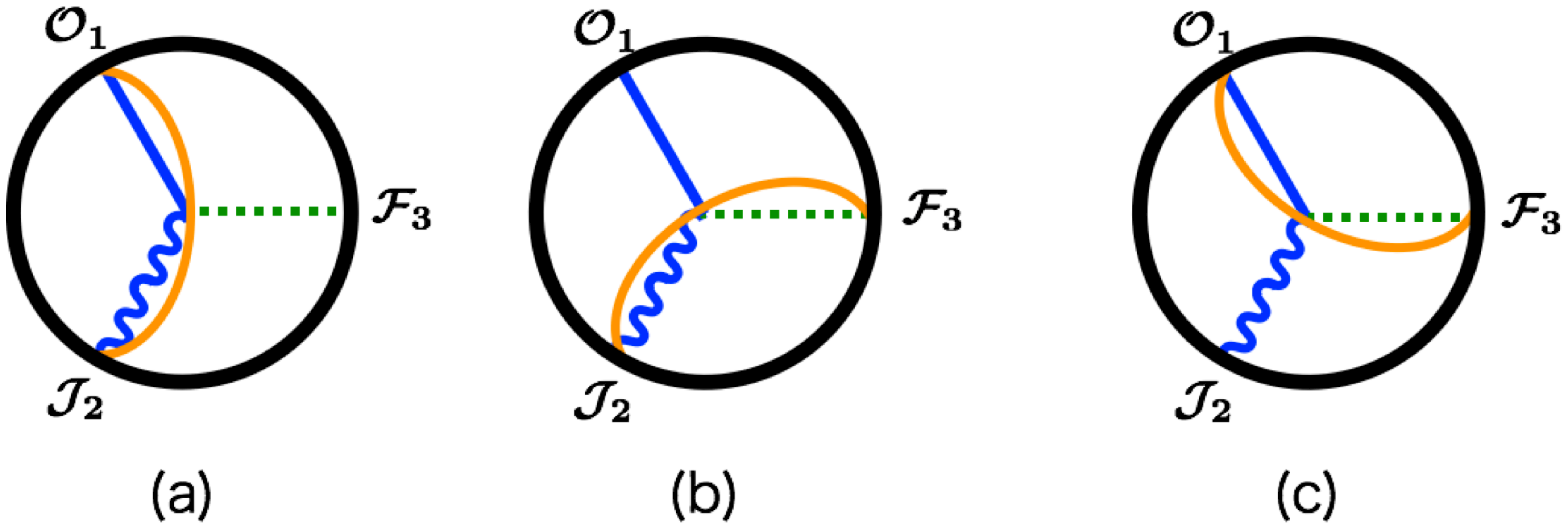}}
\caption{3pt geodesic Witten diagrams for scalar-vector-(2-form) with the different choice of the geodesics displayed as orange curves. Blue straight and wavy line represent propagation for the scalar and vector fields. Green dashed line does propagation for the 2-form field. }\label{fig:pAF}
\end{center}
\end{figure}
First of all, we estimate $\mathcal{W}_3(\gamma_{12})$ (case (a) in Figure \ref{fig:pAF}). 

After simple calculations, geodesic integral becomes
\begin{align}
\mathcal{W}_3(\gamma_{12})&=\mathcal{C}_{\mathcal{O}\mathcal{J}\mathcal{F}}\dfrac{V_{3,12}(\theta_3)H_{23}(Z_2,\theta_3)}{(-2P_{12})^{\frac{1}{2}(\D_1+\D_2-\D_3-1)}(-2P_{23})^{\frac{1}{2}(\D_2-\D_1+\D_3+3)}(-2P_{13})^{\frac{1}{2}(\D_1+\D_3-\D_2+1)}}
\end{align}
where
\begin{subequations}
\begin{align}
H_{ij}(Z_i,\theta_j)&=-2[(Z_i\cdot\theta_j)(P_i\cdot P_j)-(P_i\cdot \theta_j)(Z_i\cdot P_j)],\\
V_{i,jk}(Z_i)&=\dfrac{(Z_i\cdot P_j)(P_i\cdot P_k)-(Z_i\cdot P_k)(P_i\cdot P_j)}{(P_j\cdot P_k)},\\
V_{i,jk}(\theta_i)&=\dfrac{(\theta_i\cdot P_j)(P_i\cdot P_k)-(\theta_i\cdot P_k)(P_i\cdot P_j)}{(P_j\cdot P_k)},
\end{align}
\end{subequations}
introduced in \cite{Costa:2011mg,Costa:2014rya}. The overall coefficient is given by
\be
\mathcal{C}_{\mathcal{O}\mathcal{J}\mathcal{F}}=\la_{\phi AF}\D_1\mathcal{C}_{\D_1,0}\,\mathcal{C}_{\D_2,1}\,\mathcal{C}_{\D_3,[2]}B\left(\frac{1}{2}(\D_3+\D_1-\D_2+1)\,,\frac{1}{2}(\D_3+\D_2-\D_1+1) \right), 
\ee
where $B(x,y)$ is the beta function. 
For the computation, we used $V_{3,12}(\theta_3)V_{3,12}(\theta_3)=0$ (just the Grassmann odd property). This 3pt function agrees with the (3.57) of \cite{Costa:2014rya}. 

One can more easily check that $\mathcal{W}_3(\gamma_{23})\propto V_{3,12}(\theta_3)H_{23}(Z_2,\theta_3)$ (case (b) in Figure \ref{fig:pAF}). On the other hand, $\mathcal{W}_3(\gamma_{31})=0$ becasue $[(P_1\cdot \theta_3)(2X(\la)\cdot P_3)-(P_1\cdot P_3)(2X(\la)\cdot \theta_3)]=0$ in this case (case (c) in Figure \ref{fig:pAF}). The reader who is familiar with CFT correlators might expect any other type of 3pt interaction also reproduce the same tensor structure. Indeed, if one choose 3pt interaction as
\be
S_{int}=\la^\prime_{\phi A F}\int_{\textrm{AdS}} dX \phi_1(\p_IA_{2J})F^{IJ}_3, 
\ee
for example, we actually obtain the same function up to the overall constant. In this case, however, we observe $\mathcal{W}_3(\gamma_{23})=0$ instead. Such subtleties come from the derivative on the geodesics, which can be null. That was also observed in the case of symmetric-traceless tensors \cite{Castro:2017hpx}.  

In the same way, we can calculate, for example, 3pt GWD for vector-vector-($3$-form). If one use 
\be
S_{int}=\la_{AAF}\int_{\textrm{AdS}} dX (\p_IA_{1J})A_{2K}F^{IJK}_3,
\ee
we obtain 
$\mathcal{W}_{3}(\gamma_{12})\propto V_{3,12}(\theta_3)H_{13}(Z_1,\theta_3)H_{23}(Z_2,\theta_3)$, so does $\mathcal{W}_{3}(\gamma_{23})$. Again, $\mathcal{W}_{3}(\gamma_{13})=0$. 

\subsection{Scalar-vector-hook}
Next, we see the 3pt geodesic Witten diagram with a mixed-symmetric tensor. 
We consider a mixed-symmetric field $H$ whose irreducible representation is corresponded to the smallest hook diagram $\Yboxdim{5pt}\yng(2,1)$.\,
We consider three point interaction
\be
S_{\textrm{int.}}=\int_{AdS} dX\; (\p_I\p_K \phi_1) A_{2J} H^{IJK}_3, 
\ee
and then the amplitude is given by
\be
\mathcal{W}_3(\gamma_{ij})=\int_{\gamma_{ij}} d\la \left[(\p_I\p_KG^{\D_1,0}_{b\p}(X(\la);P_1))[G^{\D_2,1}_{b\p}(X(\la);P_2,Z_2)]_J[G^{\D_3,\,\Yboxdim{3pt}\yng(2,1)}_{b\p}(X(\la);P_3,Z_3,\theta_3)]^{IJK}\right]. 
\ee
Here we assigned\footnote{In this calculation, we are loose on the overall coefficient.}
\begin{align}
[G^{\D_3,\,\Yboxdim{3pt}\yng(2,1)}_{b\p}(X; P_3, Z_3, \theta_3)]_{IJK}&=\dfrac{1}{(-2X\cdot P_3)^{\D_3+3}}\left[2H^{Z_3}_IH^{Z_3}_JH^{\Theta_3}_K-H^{\Theta_3}_IH^{Z_3}_JH^{Z_3}_K-H^{Z_3}_IH^{\Theta_3}_JH^{Z_3}_K\right],
\end{align}
where
\begin{subequations}
\begin{align}
H^{Z_3}_A&=-2[(X\cdot P_3)Z_{3A}-(X\cdot Z_3)P_{3A}], \\
H^{\Theta_3}_A&=-2[(X\cdot P_3)\Theta_{3A}-(X\cdot \Theta_3)P_{3A}], 
\end{align}
\end{subequations}
which were discussed in section \ref{subsec:H}. This diagram turns out to be
\begin{align}
\mathcal{W}_3(\gamma_{12}), \mathcal{W}_3(\gamma_{23})&\propto\dfrac{V_{3,12}(\theta_3)V_{3,12}(Z_3)H_{23}(Z_2,Z_3)-V_{3,12}(Z_3)V_{3,12}(Z_3)H_{23}(Z_2,\theta_3)}{(-2P_{12})^{\frac{1}{2}(\D_1+\D_2-\D_3-2)}(-2P_{23})^{\frac{1}{2}(\D_2+\D_3-\D_1+4)}(-2P_{31})^{\frac{1}{2}(\D_3+\D_1-\D_2+2)}},\label{eq:svh}\\
\mathcal{W}_3(\gamma_{31})&=0. 
\end{align}
The equation \eqref{eq:svh} indeed matches the (3.62) of \cite{Costa:2014rya}. 
See also the component expression in \cite{Alkalaev:2012rg, Alkalaev:2012ic} that discussed all hook shaped diagrams. 
%%%%%%%%%%%%%%%%%%%%%%%%%%%%%%%%%%%%
\section{Conformal partial wave from geodesic Witten diagram}\label{sec:gwd}
Finally, we discuss the correspondence between GWD and CPW, both of which have the anti-symmetric exchange. We will see GWD satisfies the conformal Casimir equation and the appropriate boundary condition. We demonstrate only the simplest example, say $2$-form exchange, but extensions to more involved cases are straightforward. 
\subsection{Proof by conformal Casimir equation}
The simplest CPW with anti-symmetric exchange comes from 4pt function of two scalars and vectors: 
\be
\braket{\mathcal{O}_1(x_1)\mathcal{J}_{2a}(x_2)\mathcal{O}_3(x_3)\mathcal{J}_{4b}(x_4)}, \nn
\ee
because OPE for (scalar)\,$\times$\,(vector) can have the 2-form primary field $\mathcal{F}_{ab}$ with scaling dimension $\D$,
\begin{align}
\mathcal{O}_1(x_1)\mathcal{J}_{2a}(x_2)&\sim \left[\dfrac{C_{\footnotesize\mathcal{O}_1\mathcal{J}_2}{}^{\footnotesize\mathcal{F}}}{|x_{12}|^{\D_1+\D_2-\D+1}}x_{12}{}^{b}\mathcal{F}_{ab}(x_2)+(\textrm{decendants})\right]+(\textrm{other primaries}). 
\end{align}
Hence we have the CPW for this channel defined as follows 
\be
\left[W_{\D,[2]}(x_i)\right]_{ab}=\dfrac{1}{C_{\footnotesize\mathcal{O}_1\mathcal{J}_2}{}^{\footnotesize\mathcal{F}}C_{\footnotesize\mathcal{O}_3\mathcal{J}_4\footnotesize\mathcal{F}}}\sum_{\alpha=\mathcal{F},P\mathcal{F},\cdots}\braket{\mathcal{O}_1(x_1)\mathcal{J}_{2a}(x_2)|\alpha}\braket{\alpha|\mathcal{O}_3(x_3)\mathcal{J}_{4b}(x_4)}. \label{eq:cpw}
\ee
The explicit form of this CPW was computed in \cite{SimmonsDuffin:2012uy} by using the shadow formalism (see also \cite{Costa:2014rya}). By construction, $[W_{\D,[2]}(x_i)]_{ab}$ satisfies the conformal Casimir equation: 
\be
-\dfrac{1}{2}(L_1+L_2)^2[W_{\D,[2]}(x_i)]_{ab}=C_{\D,[2]}\,[W_{\D,[2]}(x_i)]_{ab}, 
\ee
where $L_i$ is the $SO(d+1,1)$ generators on the boundary $P_i$, and $C_{\D,[2]}=\D(\D-d)+2(d-2)$. Since three point function $\braket{\mathcal{O}_1\mathcal{J}_{a}\mathcal{F}_{bc}}$ has the unique tensor structure, we also have CPW with unique tensor structure in this channel. 
Contribution from the primary field $\mathcal{F}_{bc}$ dominates in the short distance limit $|x_{12}|\rightarrow0$, hence the behavior of CPW in this limit is $[W_{\D,[2]}]_{ab}\sim |x_{12}|^{-(\D_1+\D_2-\D+1)} x_{12}^c (\cdots)_{abc}$. Here the doted bracket $(\cdots)_{abc}$ can be read from the three point function $\braket{\mathcal{F}_{ac}\mathcal{O}_3\mathcal{J}_{4b}}$. 

\begin{figure}[tbp]
\begin{center}
\resizebox{140mm}{!}{\includegraphics{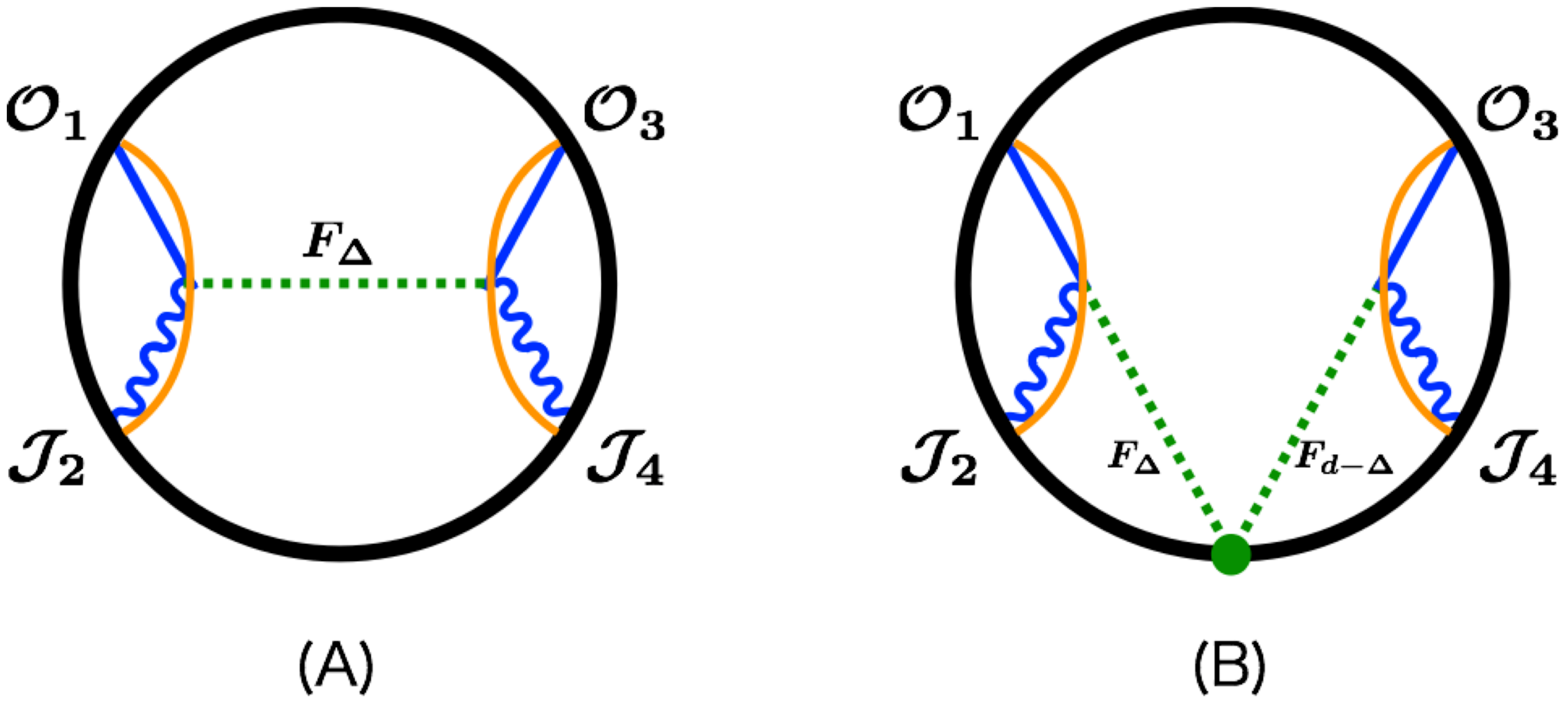}}
\caption{(A) The left side of figure shows the geodesic Witten diagram \eqref{eq:OJFOJ} which corresponds to the conformal partial wave for $\braket{\mathcal{O}_1\mathcal{J}_2\mathcal{O}_3\mathcal{J}_4}$ with $2$-form exchange \eqref{eq:cpw}. (B) The right side of figure corresponds to the bulk expression of the shadow formalism \eqref{eq:shadow}. This is schematic expression of the product of two 3pt geodesic Witten diagrams, whose common bulk point (green one) is integrated over the boundary implicitly. Two 2-form fields have scaling dimension $\D$ and $d-\D$ respectively.}\label{fig:OJFOJ}
\end{center}
\end{figure}

Let us move to the gravity dual of CPW. The corresponding amplitude of GWD (see figure \ref{fig:OJFOJ} (A)) is
\begin{align}
\mathcal{W}_{\D,[2]}(P_i;Z_2,Z_4)&=\int^{\infty}_{-\infty}\,d\la^\prime\,F[P_1,P_2,X^\prime(\la^\prime);Z_2]^{B_1B_2}(\p_{B_1}G^{\D_3,0}_{b\p}(X^\prime(\la^\prime);P_3))[G^{\D_4,1}_{b\p}(X^\prime(\la^\prime);P_4,Z_4)]_{B_2}, \label{eq:OJFOJ}
\end{align}
where we defined
\begin{align}
&\hspace{-0.5cm}F[P_1,P_2,X^\prime;Z_2]^{B_1B_2}\nn\\
&=\int^{\infty}_{-\infty}\,d\la\,[\p_{A_1}G^{\D_1,0}_{b\p}(X(\la);P_1)][G^{\D_2,1}_{b\p}(X(\la);P_2,Z_2)]_{A_2}[G^{\D,[2]}_{bb}(X(\la);X^\prime)]^{A_1A_2;B_1B_2}.
\end{align}
Here we used the geodesics
\begin{subequations}
\begin{align}
X_A(\la)&=\dfrac{P_{1A}e^{-\la}+P_{2A}e^{\la}}{\sqrt{-2P_1\cdot P_2}}, \\
X^\prime_B(\la^\prime)&=\dfrac{P_{3B}e^{-\la^\prime}+P_{4B}e^{\la^\prime}}{\sqrt{-2P_3\cdot P_4}},
\end{align}
\end{subequations}
and the component expression of the bulk-bulk propagator
\be
[G^{\D,[2]}_{bb}(X(\la);X^\prime)]_{A_1A_2;B_1B_2}=\left(\dfrac{1}{2!}\right)^2\,K^{\Theta_1}_{A_1}K^{\Theta_1}_{A_2} K^{\Theta_2}_{B_1}K^{\Theta_2}_{B_2}\,\left[G^{\D,[2]}_{bb}(X_1,\Theta_1;X_2,\Theta_2)\right]. 
\ee
In our amplitude, we chose the 3pt interaction \eqref{eq:3pint_paf} for the each interaction vertex.
The above function $F[P_1,P_2,X^\prime;Z_2]^{B_1B_2}$ is annihilated under the simultaneous rotation by $L_1+L_2+L_{X^\prime}$, where $L_{X^\prime}$ is the $SO(d+1,1)$ generator at the bulk point $X^\prime$.\footnote{Since we possess uncontracted indices $B_i$ at the point $X^\prime$, $L_{X^\prime}$ must be regarded as the generator before introducing the auxiliary field. When we consider the action onto a vector field $V_C$, for example,\be \left(L_{X^\prime}\right)_{AB}V_C=\left[X^\prime_A\dfrac{\p}{\p X^{\prime B}}-X^\prime_B\dfrac{\p}{\p X^{\prime A}}\right]V_C+(S_{AB})_{CD}V^D,\nn\ee where $(S_{AB})_{CD}=\eta_{AC}\eta_{BD}-\eta_{BC}\eta_{AD}$. }
This property leads
\begin{align}
-\dfrac{1}{2}(L_1+L_2)^2F[P_1,P_2,X^\prime;Z_2]^{B_1B_2}&=-\dfrac{1}{2}(L_{X^\prime})^2F[P_1,P_2,X^\prime;Z_2]^{B_1B_2}\\
 &=C_{\D,[2]}F[P_1,P_2,X^\prime;Z_2]^{B_1B_2}.
\end{align}
Here we used \eqref{eq:asmgwd} in the second line. Therefore, we have obtained
\be
-\dfrac{1}{2}(L_1+L_2)^2\mathcal{W}_{\D,[2]}(P_i;Z_2,Z_4)=C_{\D,[2]}\mathcal{W}_{\D,[2]}(P_i;Z_2,Z_4). 
\ee
A bit tedious but straightforward calculation shows that this integral also satisfies the desired boundary condition. To summarize, we have shown that GWD $\mathcal{W}_{\D,[2]}$ equals to CPW $W_{\D,[2]}$ up to the normalization constant. 

\subsection{Connection to the shadow formalism}\label{subsec:shadow}
It is known that the harmonic function for symmetric-traceless tensor can be defined as the product of two bulk-boundary propagators, integrated over a sharing boundary point \cite{Leonhardt:2003qu,Leonhardt:2003sn,Costa:2014kfa}. This expression is called the split representation. In particular, this expression is equivalent to the specific linear combination of bulk-bulk propagators. 

In this section, we extend this fact to the anti-symmetric tensors and see the connection between GWD and the shadow formalism\footnote{The reader who is not familiar with the shadow formalism, please see \cite{SimmonsDuffin:2012uy}.} explicitly. 
As like the symmetric-traceless case, we define the harmonic function as follows\footnote{Here the factor $(-1)^{\frac{1}{2}p(p-1)}$ is due to the left derivative of Grassmann odd variable $\theta$. },
\begin{align}
\Omega_{\nu,[p]}(X_1,\Theta_1;X_2,\Theta_2)&\equiv\dfrac{(-1)^{\frac{1}{2}p(p-1)}}{p!}\dfrac{\nu^2}{\pi}\int_{\p AdS_{d+1}}dP\; G^{h+i\nu,[p]}_{b\p}(X_1,\Theta_1;P,\p_\theta)G^{h-i\nu,[p]}_{b\p}(X_2,\Theta_2;P,\theta), \label{eq:ham}
\end{align}
where $h=d/2$. In particular, it satisfies
\begin{align}
(\nabla^2_{1}+h^2+\nu^2+p)\,\Omega_{\nu,[p]}(X_1,\Theta_1;X_2,\Theta_2)&=0, \label{eq:ham1}\\
(\nabla_{1}\cdot K^{\Theta_1})\,\Omega_{\nu,[p]}(X_1,\Theta_1;X_2,\Theta_2)&=0.\label{eq:ham2}
\end{align}
Interestingly, the above definition \eqref{eq:ham} is equivalent to
\begin{align}
\Omega_{\nu,[p]}(X_1,\Theta_1;X_2,\Theta_2)&=\dfrac{i\nu}{2\pi}\left(G^{h+i\nu,[p]}_{bb}(X_1,\Theta_1;X_2,\Theta_2)-G^{h-i\nu,[p]}_{bb}(X_1,\Theta_1;X_2,\Theta_2)\right). \label{eq:hambb}
\end{align}
We note the derivation of \eqref{eq:hambb} in appendix \ref{app:ham}. 
Notice that $h-i\nu=d-(h+i\nu)$. Therefore this relation suggests the bulk expression of the shadow formalism. Namely, there are also the contribution from the shadow operator with the scaling dimension $d-\D$ to which we want project out. Replacing our bulk-bulk propagator to $\Omega_{-i(\D-h),[p]}$, we obtain
\begin{align}
\widetilde{\mathcal{W}}_{\mathcal{F}}&=\int_{\gamma_{12}} d\la\int_{\gamma_{34}} d\la^\prime \Bigg\{\Big[(K^{\Theta_1}\cdot\p_X)G^{\D_1,0}_{b\p}(X(\la);P_1)\Big]\Big[G^{\D_2,1}_{b\p}(X(\la),K^{\Theta_1};P_2,Z_2)\Big]\nn\\
&\hspace{4cm}\times\Omega_{-i(\D-h),[p]}(X(\la),\Theta_1;X^\prime(\la^\prime),K^{\Theta_2})\nn\\
&\hspace{4cm}\times\Big[(\Theta_2\cdot\p_{X^\prime})G^{\D_3,0}_{b\p}(X^\prime(\la^\prime),P_3)\Big]\Big[G^{\D_4,1}_{b\p}(X^\prime(\la^\prime),\Theta_2;P_4,Z_4)\Big]\Bigg\}\label{eq:shadow}
\end{align}
Here contraction for indices are represented as the differential operator $K^{\Theta}$ introduced in \eqref{eq:K}.
This expression equals the product of two 3pt GWD discussed in section \ref{subsec:paf}, integrated over a sharing boundary point (see figure \ref{fig:OJFOJ} (B)). Since these 3pt GWD are equivalent to the 3pt conformal correlators, \eqref{eq:shadow} is nothing but the bulk expression of the shadow formalism,
\be
\widetilde{\mathcal{W}}_{\mathcal{F}}\propto\int_{\p AdS_{d+1}}dP\;\braket{\mathcal{O}_1(P_1)\mathcal{J}_{2}(P_2,Z_2)(\mathcal{F}_\D(P))_{cd}}\braket{(\mathcal{F}_{d-\D}(P))^{cd}\mathcal{O}_3(P_3)\mathcal{J}_{4}(P_4,Z_4)}.
\ee
Inverting the above argument, we can confirm that the bulk expression naturally distinguishes between the desired CPW and its shadow contribution. 
\section{Summary and Discussion}\label{sec:summary}
In this paper, we have developed the embedding formalism for anti-symmetric fields in AdS and applied it to the GWD with an anti-symmetric field. 
We have seen that 3pt GWD including the anti-symmetric field reproduce 3pt CFT correlators up to constant. We have also extended such computation to the 3pt GWD with a mixed-symmetric tensor and obtained the corresponding one. There are a number of choices of the 3pt interaction for these diagrams, but some of them vanish. This is the subtlety of the geodesic integrals and we need to take care about it, for example, when decomposing the usual Witten diagram into the sum of the GWD.  

Moreover, we have shown that 4pt GWD with the anti-symmetric exchange also reproduces the CPW with the anti-symmetric primary field. The point was that the free equation of motions in AdS${}_{d+1}$ equals the conformal Casimir equation in CFT${}_d$. Moreover, we have discussed the split representation of the harmonic function for anti-symmetric tensors. This manifests the connection between GWD and the shadow formalism. We can expect that these properties do hold for any types of representations, such as mixed-symmetric tensors. However, there are no explicit form of the bulk-bulk propagators with mixed-symmetric tensors in the literatures. It might be interesting to find the explicit form of bulk-bulk propagators of interests and its split representation.

Throughout this paper, we yielded the case when corresponding CFT correlators have the unique tensor structure just for simplicity. For the extension to the more involved cases with non-unique tensor structures, we can apply the differential operators \cite{Costa:2011dw} for the symmetric indices as like previous works. Thus, for external fields with mixed-symmetry, it might be useful to find the ones for mixed-symmetric indices.

It would be also interesting to develop the embedding formalism in AdS for fields including the spinor indices, explore the Witten diagram on the minimal surface \cite{Czech:2016xec}, and find the relation among the seed geodesic Witten diagrams \cite{Karateev:2017jgd}. We leave these questions for the future work.
%\if(
\bigskip
\goodbreak
\centerline{\bf Acknowledgments}
\noindent
We are grateful to Mitsuhiro~Nishida, Satoshi~Yamaguchi, and especially Norihiro~Iizuka for valuable comments, discussion, and encouragement. 
%)\fi
%\newpage
\appendix
\section{Split representation of the bulk-bulk propagators}\label{app:ham}
In this appendix, we derive some properties of the AdS harmonic function for $p$-form fields. The definition of harmonic function is given by
\begin{align}
\Omega_{\nu,[p]}(X_1,\Theta_1;X_2,\Theta_2)&\equiv(-1)^{\frac{1}{2}p(p-1)}\dfrac{\nu^2}{p!\pi}\int_{\p AdS_{d+1}}dP\;G^{h+i\nu,[p]}_{b\p}(X_1,\Theta_1;P,\p_\theta)G^{h-i\nu,[p]}_{b\p}(X_2,\Theta_2;P,\theta). \label{eq:appham}
\end{align}
Since $\Omega_{\nu,[p]}$ consists of bulk-boundary propagators for $p$-form, the properties \eqref{eq:ham1} and \eqref{eq:ham2} readily follow from the properties of bulk-boundary propagators. 
Thus we call \eqref{eq:appham} harmonic function. 

Next, we see the definition \eqref{eq:appham} is equivalent to the difference of bulk-bulk propagators
\be
\Omega_{\nu,[p]}(X_1,\Theta_1;X_2,\Theta_2)=\dfrac{i\nu}{2\pi}\Big[G^{h+i\nu,[p]}_{bb}(X_1,\Theta_1;X_2,\Theta_2)-G^{h-i\nu,[p]}_{bb}(X_1,\Theta_1;X_2,\Theta_2)\Big].\label{eq:appbb} 
\ee
The tensor part of \eqref{eq:appham} is
\begin{align}
&[-2(\Theta_1\cdot\p_\theta)(X_1\cdot P)+2(\Theta_1\cdot P)(X_1\cdot \p_\theta)]^p[-2(\Theta_2\cdot\theta)(X_2\cdot P)+2(\Theta_2\cdot P)(X_2\cdot\theta)]^p\label{eq:hamnu}
\end{align}
We can choose polarizations such that $\Theta_i\cdot X_j=0$. This specific choice simplifies the \eqref{eq:hamnu} as follows:
\begin{align}
&(-1)^{\frac{1}{2}p(p-1)}\,p!\,\left[(-2X_1\cdot P)^p(-2X_2\cdot P)^p(\Theta_1\cdot\Theta_2)^p+4(X_1\cdot X_2)(\Theta_1\cdot\Theta_2)^{p-1}(\Theta_1\cdot P)(\Theta_2\cdot P)\right]
\end{align}
Then integration for the boundary point $P$ can be done by using Feynman parametrization
\be
\frac{1}{A^xB^y}=\frac{\G(x+y)}{\G(x)\G(y)}\int^\infty_0\frac{dt}{t}t^y\frac{1}{[A+tB]^{x+y}},
\ee
and the formula \cite{SimmonsDuffin:2012uy}
\be
\int_{\p AdS_{d+1}}dP\;\dfrac{P_AP_B}{[-2X\cdot P]^{2h+2}}=\dfrac{\pi^h\Gamma(h+2)}{\Gamma(2h+2)}\dfrac{1}{(-X^2)^{h+2}}\left[X_AX_B-\dfrac{1}{2h+2}\eta_{AB}X^2\right].
\ee
Remaining integral for Feynman parameter $t$ can be finished by using the equality
\begin{align}
\int^\infty_0\dfrac{dt}{t}\dfrac{t^{-c}}{\left(2U+\frac{(1+t)^2}{t}\right)^b}&=\dfrac{\Gamma(b+c)\Gamma(-c)}{\Gamma(b)(2U)^{b+c}}{}_2F_1\left(b+c,c+\frac{1}{2},2c+1;-\frac{2}{U}\right)\nn\\
&+\dfrac{\Gamma(b-c)\Gamma(c)}{\Gamma(b)(2U)^{b-c}}{}_2F_1\left(b-c,-c+\frac{1}{2},-2c+1;-\frac{2}{U}\right). 
\end{align}
The result which is proportional to $(\Theta_1\cdot\Theta_2)^p$ turns out to be the difference between $g_0(U)$ in \eqref{eq:g0} with scaling dimension $h+i\nu$ and one with $h-i\nu$ (and numerical prefactor $\frac{i\nu}{2\pi}$). 
Therefore we have shown that the definition \eqref{eq:appham} and \eqref{eq:appbb} is equivalent.


\begin{thebibliography}{99}
%\cite{Polyakov:1970xd}
\bibitem{Polyakov:1970xd} 
  A.~M.~Polyakov,
  ``Conformal symmetry of critical fluctuations,''
  JETP Lett.\  {\bf 12}, 381 (1970)
  [Pisma Zh.\ Eksp.\ Teor.\ Fiz.\  {\bf 12}, 538 (1970)].

%\cite{Belavin:1984vu}
\bibitem{Belavin:1984vu} 
  A.~A.~Belavin, A.~M.~Polyakov and A.~B.~Zamolodchikov,
  ``Infinite Conformal Symmetry in Two-Dimensional Quantum Field Theory,''
  Nucl.\ Phys.\ B {\bf 241}, 333 (1984).

%\cite{Maldacena:1997re}
\bibitem{Maldacena:1997re} 
  J.~M.~Maldacena,
  ``The Large N limit of superconformal field theories and supergravity,''
  Int.\ J.\ Theor.\ Phys.\  {\bf 38}, 1113 (1999)
  [Adv.\ Theor.\ Math.\ Phys.\  {\bf 2}, 231 (1998)]
  %doi:10.1023/A:1026654312961
  [hep-th/9711200].

  %\cite{Gubser:1998bc}
\bibitem{Gubser:1998bc} 
  S.~S.~Gubser, I.~R.~Klebanov and A.~M.~Polyakov,
  ``Gauge theory correlators from noncritical string theory,''
  Phys.\ Lett.\ B {\bf 428}, 105 (1998)
  %doi:10.1016/S0370-2693(98)00377-3
  [hep-th/9802109].
  
  %\cite{Witten:1998qj}
\bibitem{Witten:1998qj} 
  E.~Witten,
  ``Anti-de Sitter space and holography,''
  Adv.\ Theor.\ Math.\ Phys.\  {\bf 2}, 253 (1998)
  [hep-th/9802150].

  %\cite{Rattazzi:2008pe}
\bibitem{Rattazzi:2008pe} 
  R.~Rattazzi, V.~S.~Rychkov, E.~Tonni and A.~Vichi,
  ``Bounding scalar operator dimensions in 4D CFT,''
  JHEP {\bf 0812}, 031 (2008)
  %doi:10.1088/1126-6708/2008/12/031
  [arXiv:0807.0004 [hep-th]].
  
  %\cite{ElShowk:2012ht}
\bibitem{ElShowk:2012ht} 
  S.~El-Showk, M.~F.~Paulos, D.~Poland, S.~Rychkov, D.~Simmons-Duffin and A.~Vichi,
  ``Solving the 3D Ising Model with the Conformal Bootstrap,''
  Phys.\ Rev.\ D {\bf 86}, 025022 (2012)
  %doi:10.1103/PhysRevD.86.025022
  [arXiv:1203.6064 [hep-th]].
  
%\cite{Hijano:2015zsa}
\bibitem{Hijano:2015zsa} 
  E.~Hijano, P.~Kraus, E.~Perlmutter and R.~Snively,
  ``Witten Diagrams Revisited: The AdS Geometry of Conformal Blocks,''
  JHEP {\bf 1601}, 146 (2016)
  %doi:10.1007/JHEP01(2016)146
  [arXiv:1508.00501 [hep-th]].

%\cite{Heemskerk:2009pn}
\bibitem{Heemskerk:2009pn} 
  I.~Heemskerk, J.~Penedones, J.~Polchinski and J.~Sully,
  ``Holography from Conformal Field Theory,''
  JHEP {\bf 0910}, 079 (2009)
  %doi:10.1088/1126-6708/2009/10/079
  [arXiv:0907.0151 [hep-th]].

%\cite{ElShowk:2011ag}
\bibitem{ElShowk:2011ag} 
  S.~El-Showk and K.~Papadodimas,
  ``Emergent Spacetime and Holographic CFTs,''
  JHEP {\bf 1210}, 106 (2012)
  %doi:10.1007/JHEP10(2012)106
  [arXiv:1101.4163 [hep-th]].

%\cite{Dolan:2003hv}
\bibitem{Dolan:2003hv} 
  F.~A.~Dolan and H.~Osborn,
  ``Conformal partial waves and the operator product expansion,''
  Nucl.\ Phys.\ B {\bf 678}, 491 (2004)
  %doi:10.1016/j.nuclphysb.2003.11.016
  [hep-th/0309180].

    %\cite{Ferrara:1972xe}
\bibitem{Ferrara:1972xe} 
  S.~Ferrara and G.~Parisi,
  ``Conformal covariant correlation functions,''
  Nucl.\ Phys.\ B {\bf 42}, 281 (1972).

  %\cite{Ferrara:1972uq}
\bibitem{Ferrara:1972uq} 
  S.~Ferrara, A.~F.~Grillo, G.~Parisi and R.~Gatto,
  ``The shadow operator formalism for conformal algebra. vacuum expectation values and operator products,''
  Lett.\ Nuovo Cim.\  {\bf 4S2}, 115 (1972)

  %\cite{Ferrara:1972ay}
\bibitem{Ferrara:1972ay} 
  S.~Ferrara, A.~F.~Grillo and G.~Parisi,
  ``Nonequivalence between conformal covariant wilson expansion in euclidean and minkowski space,''
  Lett.\ Nuovo Cim.\  {\bf 5S2}, 147 (1972)

%\cite{Ferrara:1973vz}
\bibitem{Ferrara:1973vz} 
  S.~Ferrara, A.~F.~Grillo, G.~Parisi and R.~Gatto,
  ``Covariant expansion of the conformal four-point function,''
  Nucl.\ Phys.\ B {\bf 49}, 77 (1972)
  Erratum: [Nucl.\ Phys.\ B {\bf 53}, 643 (1973)].

  %\cite{Ferrara:1974nf}
\bibitem{Ferrara:1974nf} 
  S.~Ferrara, A.~F.~Grillo, R.~Gatto and G.~Parisi,
  ``Analyticity properties and asymptotic expansions of conformal covariant green's functions,''
  Nuovo Cim.\ A {\bf 19}, 667 (1974).
  
%\cite{SimmonsDuffin:2012uy}
\bibitem{SimmonsDuffin:2012uy} 
  D.~Simmons-Duffin,
  ``Projectors, Shadows, and Conformal Blocks,''
  JHEP {\bf 1404}, 146 (2014)
  %doi:10.1007/JHEP04(2014)146
  [arXiv:1204.3894 [hep-th]].

%\cite{Vasiliev:2004cm}
\bibitem{Vasiliev:2004cm} 
  M.~A.~Vasiliev,
  ``Higher spin superalgebras in any dimension and their representations,''
  JHEP {\bf 0412}, 046 (2004)
  %doi:10.1088/1126-6708/2004/12/046
  [hep-th/0404124].

%\cite{Nishida:2016vds}
\bibitem{Nishida:2016vds} 
  M.~Nishida and K.~Tamaoka,
  ``Geodesic Witten diagrams with an external spinning field,''
  arXiv:1609.04563 [hep-th].

%\cite{Castro:2017hpx}
\bibitem{Castro:2017hpx} 
  A.~Castro, E.~Llabres and F.~Rejon-Barrera,
  ``Geodesic Diagrams, Gravitational Interactions \& OPE Structures,''
  arXiv:1702.06128 [hep-th].

%\cite{Dyer:2017zef}
\bibitem{Dyer:2017zef} 
  E.~Dyer, D.~Z.~Freedman and J.~Sully,
  ``Spinning Geodesic Witten Diagrams,''
  arXiv:1702.06139 [hep-th].

%\cite{Sleight:2017fpc}
\bibitem{Sleight:2017fpc} 
  C.~Sleight and M.~Taronna,
  ``Spinning Witten Diagrams,''
  JHEP {\bf 1706}, 100 (2017)
  %doi:10.1007/JHEP06(2017)100
  [arXiv:1702.08619 [hep-th]].

%\cite{Chen:2017yia}
\bibitem{Chen:2017yia} 
  H.~Y.~Chen, E.~J.~Kuo and H.~Kyono,
  ``Anatomy of Geodesic Witten Diagrams,''
  JHEP {\bf 1705}, 070 (2017)
  %doi:10.1007/JHEP05(2017)070
  [arXiv:1702.08818 [hep-th]].

%\cite{Rastelli:2017ecj}
\bibitem{Rastelli:2017ecj} 
  L.~Rastelli and X.~Zhou,
  ``The Mellin Formalism for Boundary CFT$_d$,''
  arXiv:1705.05362 [hep-th].

%\cite{Bena:1999be}
\bibitem{Bena:1999be} 
  I.~Bena,
  ``The Propagator for a general form field in AdS(d+1),''
  Phys.\ Rev.\ D {\bf 62}, 126008 (2000)
  %doi:10.1103/PhysRevD.62.126008
  [hep-th/9911073].

%\cite{Bena:1999py}
\bibitem{Bena:1999py} 
  I.~Bena,
  ``The Antisymmetric tensor propagator in AdS,''
  Phys.\ Rev.\ D {\bf 62}, 127901 (2000)
  %doi:10.1103/PhysRevD.62.127901
  [hep-th/9910059].

%\cite{Naqvi:1999va}
\bibitem{Naqvi:1999va} 
  A.~Naqvi,
  ``Propagators for massive symmetric tensor and p forms in AdS(d+1),''
  JHEP {\bf 9912}, 025 (1999)
  %doi:10.1088/1126-6708/1999/12/025
  [hep-th/9911182].

%\cite{Costa:2014kfa}
\bibitem{Costa:2014kfa} 
  M.~S.~Costa, V.~Goncalves and J.~Penedones,
  ``Spinning AdS Propagators,''
  JHEP {\bf 1409}, 064 (2014)
  %doi:10.1007/JHEP09(2014)064
  [arXiv:1404.5625 [hep-th]].

%\cite{Joung:2011ww}
\bibitem{Joung:2011ww} 
  E.~Joung and M.~Taronna,
  ``Cubic interactions of massless higher spins in (A)dS: metric-like approach,''
  Nucl.\ Phys.\ B {\bf 861}, 145 (2012)
  %doi:10.1016/j.nuclphysb.2012.03.013
  [arXiv:1110.5918 [hep-th]].

%\cite{Costa:2011mg}
\bibitem{Costa:2011mg} 
  M.~S.~Costa, J.~Penedones, D.~Poland and S.~Rychkov,
  ``Spinning Conformal Correlators,''
  JHEP {\bf 1111}, 071 (2011)
  %doi:10.1007/JHEP11(2011)071
  [arXiv:1107.3554 [hep-th]].

%\cite{Costa:2011dw}
\bibitem{Costa:2011dw} 
  M.~S.~Costa, J.~Penedones, D.~Poland and S.~Rychkov,
  ``Spinning Conformal Blocks,''
  JHEP {\bf 1111}, 154 (2011)
  %doi:10.1007/JHEP11(2011)154
  [arXiv:1109.6321 [hep-th]].

%\cite{Costa:2014rya}
\bibitem{Costa:2014rya}
  M.~S.~Costa and T.~Hansen,
  ``Conformal correlators of mixed-symmetry tensors,''
  JHEP {\bf 1502} (2015) 151
  %doi:10.1007/JHEP02(2015)151
  [arXiv:1411.7351 [hep-th]].
  
%EMBEDDING FORMALISM%
%\cite{Dirac:1936fq}
\bibitem{Dirac:1936fq} 
  P.~A.~M.~Dirac,
  ``Wave equations in conformal space,''
  Annals Math.\  {\bf 37}, 429 (1936).

%\cite{Mack:1969rr}
\bibitem{Mack:1969rr} 
  G.~Mack and A.~Salam,
  ``Finite component field representations of the conformal group,''
  Annals Phys.\  {\bf 53}, 174 (1969).
  
%\cite{Boulware:1970ty}
\bibitem{Boulware:1970ty} 
  D.~G.~Boulware, L.~S.~Brown and R.~D.~Peccei,
  ``Deep-inelastic electroproduction and conformal symmetry,''
  Phys.\ Rev.\ D {\bf 2}, 293 (1970).

%\cite{Ferrara:1973eg}
\bibitem{Ferrara:1973eg} 
  S.~Ferrara, R.~Gatto and A.~F.~Grillo,
  ``Conformal algebra in space-time and operator product expansion,''
  Springer Tracts Mod.\ Phys.\  {\bf 67}, 1 (1973).

%\cite{Ferrara:1973yt}
\bibitem{Ferrara:1973yt} 
  S.~Ferrara, A.~F.~Grillo and R.~Gatto,
  ``Tensor representations of conformal algebra and conformally covariant operator product expansion,''
  Annals Phys.\  {\bf 76}, 161 (1973).

%\cite{Cornalba:2009ax}
\bibitem{Cornalba:2009ax} 
  L.~Cornalba, M.~S.~Costa and J.~Penedones,
  ``Deep Inelastic Scattering in Conformal QCD,''
  JHEP {\bf 1003}, 133 (2010)
  %doi:10.1007/JHEP03(2010)133
  [arXiv:0911.0043 [hep-th]].

%\cite{Weinberg:2010fx}
\bibitem{Weinberg:2010fx} 
  S.~Weinberg,
  ``Six-dimensional Methods for Four-dimensional Conformal Field Theories,''
  Phys.\ Rev.\ D {\bf 82}, 045031 (2010)
  %doi:10.1103/PhysRevD.82.045031
  [arXiv:1006.3480 [hep-th]].

%\cite{Costa:2016hju}
\bibitem{Costa:2016hju} 
  M.~S.~Costa, T.~Hansen, J.~Penedones and E.~Trevisani,
  ``Projectors and seed conformal blocks for traceless mixed-symmetry tensors,''
  JHEP {\bf 1607}, 018 (2016)
  %doi:10.1007/JHEP07(2016)018
  [arXiv:1603.05551 [hep-th]].

%\cite{Alkalaev:2012rg}
\bibitem{Alkalaev:2012rg} 
  K.~Alkalaev,
  ``Mixed-symmetry tensor conserved currents and AdS/CFT correspondence,''
  J.\ Phys.\ A {\bf 46}, 214007 (2013)
  %doi:10.1088/1751-8113/46/21/214007
  [arXiv:1207.1079 [hep-th]].
  
%\cite{Alkalaev:2012ic}
\bibitem{Alkalaev:2012ic} 
  K.~Alkalaev,
  ``Massless hook field in AdS(d+1) from the holographic perspective,''
  JHEP {\bf 1301}, 018 (2013)
  %doi:10.1007/JHEP01(2013)018
  [arXiv:1210.0217 [hep-th]].
  
%\cite{Leonhardt:2003qu}
\bibitem{Leonhardt:2003qu}
  T.~Leonhardt, R.~Manvelyan and W.~Ruhl,
  ``The Group approach to AdS space propagators,''
  Nucl.\ Phys.\ B {\bf 667} (2003) 413
  %doi:10.1016/j.nuclphysb.2003.07.007
  [hep-th/0305235].
  
%\cite{Leonhardt:2003sn}
\bibitem{Leonhardt:2003sn} 
  T.~Leonhardt, W.~Ruhl and R.~Manvelyan,
  ``The Group approach to AdS space propagators: A Fast algorithm,''
  J.\ Phys.\ A {\bf 37}, 7051 (2004)
  %doi:10.1088/0305-4470/37/27/013
  [hep-th/0310063].

%\cite{Czech:2016xec}
\bibitem{Czech:2016xec} 
  B.~Czech, L.~Lamprou, S.~McCandlish, B.~Mosk and J.~Sully,
  ``A Stereoscopic Look into the Bulk,''
  JHEP {\bf 1607}, 129 (2016)
 % doi:10.1007/JHEP07(2016)129
  [arXiv:1604.03110 [hep-th]].

%\cite{Karateev:2017jgd}
\bibitem{Karateev:2017jgd} 
  D.~Karateev, P.~Kravchuk and D.~Simmons-Duffin,
  ``Weight Shifting Operators and Conformal Blocks,''
  arXiv:1706.07813 [hep-th].
\end{thebibliography}
\end{document}